\documentstyle[preprint,aps,floats]{revtex}
\tightenlines
\begin{document}
\def\ba{\begin{eqnarray}}
\def\ea{\end{eqnarray}}
\def\be{\begin{equation}}
\def\ee{\end{equation}}
\def\t{\tau}

\title{Living with Ghosts}

\author{S.W. Hawking\thanks{S.W.Hawking@damtp.cam.ac.uk},
Thomas Hertog\thanks{T.Hertog@damtp.cam.ac.uk}
 \\ $\ $ \\
DAMTP \\
Centre for Mathematical Sciences \\
Wilberforce Road, Cambridge, CB3 0WA, UK.}
\date{\today}
\maketitle

\begin{abstract}

Perturbation theory for gravity in dimensions greater than two requires 
higher derivatives in the free action.
Higher derivatives seem to lead to ghosts, states with negative norm.
We consider a fourth order scalar field theory
and show that the problem with ghosts arises because in the canonical 
treatment, $\phi$ and $\Box \phi $ are regarded as two independent variables.
Instead, we base quantum theory on a path integral, 
evaluated in Euclidean space and then Wick rotated to Lorentzian 
space.
The path integral requires that quantum states be specified by the values of 
$\phi$ and $\phi_{,\tau}$.
To calculate probabilities for observations, one has to trace out over
$\phi_{,\tau}$ on the final surface.
Hence one loses unitarity, but one can never produce a 
negative norm state or get a negative probability.   
It is shown that transition probabilities tend toward those of the
second order theory, as the coefficient of the fourth order term in the 
action tends to zero. Hence unitarity is restored at the low energies 
that now occur in the universe.

\end{abstract}
\vskip .2in

\section{Introduction}

In standard, second order theory the Lagrangian is a function of the fields 
and their first derivatives. 
The path integral is calculated by perturbation theory, with
the part of the action that contains quadratic terms in the fields and 
their first derivatives regarded as the free field action, and  
the remaining terms as interactions. One then 
calculates Feynman diagrams, using the interactions as vertices, and the 
propagator defined by the free part of the action.  This is
equivalent to calculating the expectation value of the interactions in the 
Gaussian measure defined by the free action.  One would therefore expect 
perturbation theory to make sense, when and only when, the interaction action
is bounded by the free action. 


This is born out by the examples we know.
In two dimensions, the free action of a scalar field $\phi$, 
\be
S = \int dx^2 \left[ \phi \Box \phi +m^2 \phi^2 \right],
\ee
is the first Sobolev norm\footnote{For 
a function $f \in C^{\infty}(M)$, $1 \leq p < \infty $, and an integer
$k \geq 0$, the Sobolev norm is defined \cite{Besse87} as
\be
\Vert f \Vert_{p,k} = \left[ \int_{M} 
\sum_{0 \leq j \leq k} \vert D^{j} f \vert^{p}  \mu_{g} \right]^{1/p},
\ee
where $\vert D^{j} f \vert $ is the pointwise norm of the jth
covariant derivative and $\mu_{g}$ is the Riemannian volume element.}
$\Vert \phi \Vert_{2,1}$ of the field $\phi$. 
In two dimensions, the first Sobolev norm bounds the pointwise value of 
$\phi$, thus it also bounds the volume integral of any entire 
function of $\phi$. This means that the free action bounds any interaction 
action, so perturbation theory should work. Indeed one finds that in two 
dimensions, any quantum field theory is renormalizable.

In four dimensions on the other hand, the first Sobolev norm does not bound 
the pointwise value of $\phi$, but only 
the volume integral of $\phi^4$. 
This means that the free action bounds the interactions only for theories 
with quartic interactions, like $\lambda \phi^4$, or Yang--Mills. 
Indeed, these are the quantum field theories that are renormalizable in four 
dimensions. Note that even Yang--Mills is not renormalizable in dimensions 
higher than four, because the interactions are not bounded by the free 
action. Similarly, Born--Infeld is not renormalizable in dimensions higher 
than two. 

When one does perturbation theory for gravity, one writes the 
metric as $g_0 + \delta g$, where $g_0$ is a background metric that is a 
solution of the field equations. 
The terms quadratic in $\delta g$ are again regarded as the free action, 
and the higher order terms are the interactions. The latter
include terms like $(\nabla \delta g)^2$, multiplied by powers of $\delta g$.
The volume integral of such an interaction is not bounded by the free 
action and perturbation theory breaks down for gravity, which is not 
renormalizable \cite{thooft}. 
Even if all the higher loop divergences canceled by some 
miracle in a supergravity theory, one couldn't 
trust the results, because one is using perturbation theory beyond its limit 
of validity; $\delta g$ can be much larger than $g_0$ locally for only a 
small free action. 
In other words, there are large metric fluctuations below the 
Planck scale. 

The situation is different however if one adds curvature 
squared terms to the Einstein--Hilbert action. The action is now 
quadratic in second derivatives of $\delta g$, so 
one takes the free action to be the quadratic terms in $\delta g$, 
and its first and second derivatives. 
This means that it is the second Sobolev norm $\Vert \delta g
\Vert_{2,2}$ of $\delta g$, which
bounds the pointwise value of $\delta g$. Hence
the free action bounds the interactions, and perturbation theory works. 
This is reflected in the fact that the $R+R^2$ theory is renormalizable
\cite{Stelle}, and in fact asymptotically free \cite{Fradkin}. 
However, higher derivatives seem to lead to ghosts, 
states with negative norm, which 
have been thought to be a fatal flaw in any quantum field theory (see e.g.
\cite{Hawking85}). 

In the next section we review why higher derivatives appear to 
give rise to ghosts. The existence of ghosts 
would mean that the set of all states would not form a Hilbert space with a 
positive definite metric. There would not be a unitary S matrix, and there 
would apparently be states with negative probabilities. These seemed 
sufficient reasons to dismiss any quantum field theory, such as Einstein 
gravity, that had higher derivative quantum corrections and ghosts. 
However, we shall show that one can still make sense of higher 
derivative theories, as a set of rules for calculating probabilities for 
observations. But one can 
not prepare a system in a state with a negative norm, nor can one resolve a 
state into its positive and negative norm components.  So there are no 
negative probabilities, and no non unitary S matrix. 

Although gravity is the physically interesting case, in this paper 
we consider a fourth order scalar field theory,
which has the same ghostly behaviour, but
doesn't have the complications of indices or gauge invariance. 
We show explicitely that the higher derivative theory  
tends toward the second order theory, as the coefficient of the 
fourth order term in the action tends to zero.
Hence the departures from unitarity for higher derivative gravity are very 
small at the low energies that now occur in the universe.


\section{Higher Derivative Ghosts}


We consider a scalar field $\phi$ with a 
fourth-order Lagrangian in Lorentzian signature, 

\be \label{fourth}
L = -\frac{1}{2}\phi \left( \Box -m_{1}^{2} \right)
 \left( \Box -m_{2}^{2} \right)\phi - \lambda \phi^4
\ee
where $m_2 > m_1$.
Defining
\be \label{ps}
\psi_1 = \frac{ \left( \Box -m_{2}^{2} \right) \phi}{ [2(m_{2}^{2}
-m_{1}^{2})]^{1/2}}
\qquad 
\qquad
\psi_2 = \frac{ \left( \Box -m_{1}^{2} \right) \phi}{ [2(m_{2}^{2}
-m_{1}^{2})]^{1/2}}
\ee
the Lagrangian can be rewritten as
\be
L = \frac{1}{2} \psi_1 \left( \Box -m_{1}^{2} \right)\psi_1
-\frac{1}{2} \psi_2 \left( \Box -m_{2}^{2} \right)\psi_2
-\frac{4 \lambda}{(m^{2}_{2}-m^{2}_{1})^2}(\psi_1 - \psi_2 )^4
\ee 

The action of $\psi_2$ has the wrong sign. Classically it means that 
the energy of the $\psi_2$ field is negative, while that of $\psi_1$ is 
positive. If there were no interaction term, this negative energy wouldn't 
matter because each of the fields, $\psi_1$ and $\psi_2$, would live in its 
own world and the two worlds would not communicate with each other. 
However, if there is an interaction term, like $\phi^4$, it will couple 
$\psi_1$ and 
$\psi_2$ together. Energy can then flow from one to the other, 
and one can have  
runaway solutions, with the positive energy of $\psi_1$ and the negative 
energy of $\psi_2$ both increasing exponentially.

In quantum theory, on the other hand, one is in trouble even in the absence 
of interactions, as can be seen by looking at the free field propagator for 
$\phi$. In momentum space, this is the inverse of a fourth order expression 
in $p$, which can be expanded as 
\be
G(p) =
\frac{1}{(m^{2}_{2}-m^{2}_{1})}
\left( \frac{1}{(p^2 +m_{1}^{2})} -\frac{1}{(p^2 +m_{2}^{2})} \right),
\ee
This is just the difference of the propagators for $\psi_1$ and $\psi_2$. 
The important point is that the propagator for $\psi_2$ 
appears with a negative sign. This would mean that states with an odd number 
of $\psi_2$ particles, would have a negative norm. In other words, $\psi_2$  
particles are ghosts. There wouldn't be a positive definite Hilbert space 
metric, nor a unitary S matrix.

If there weren't any interactions, the situation wouldn't be too serious. 
The state space would be the direct sum of two Hilbert spaces, one with 
positive definite metric and the other negative. There wouldn't be any 
physically realized operators that connected the two Hilbert spaces, so 
ghost number would be conserved by a superselection rule. A $\phi^4$ 
interaction however, would allow $\psi_2$ particles to be created or 
destroyed. As in the classical theory, there will be instabilities, with 
runaway production of $\psi_1$ and $\psi_2$ particles. 
These instabilities show up in the fact that interactions tend to shift the 
ghost poles in the two point 
function for $\phi$ into the complex $p$-plane, where they represent 
exponentially growing and decaying modes \cite{Tomboulis77,Johnston87}.

It seems to add up to a 
pretty damning indictment of higher derivative theories in general, and 
quantum gravity and quantum supergravity in particular.  
However, the problem with ghosts arises because in the canonical treatment, 
$\phi$ and $\Box \phi$ are regarded as two independent variables, 
although they are both determined by $\phi$.
We shall show that, by basing quantum theory on a path integral
over the field, 
evaluated in Euclidean space and then Wick rotated to Lorentzian space,
one can obtain a sensible set of rules for calculating probabilities 
for observations in higher derivative theories.

\section{Euclidean Path Integral}

According to the canonical approach, one would perform the path 
integral over all $\psi_1$ and $\psi_2$. 
The path integral over $\psi_1$ will converge, but the path integral over 
$\psi_2$ is divergent, because the free action for $\psi_2$ is negative 
definite. 
However, one shouldn't do the path integrals over $\psi_1$ and $\psi_2$ 
separately because they are not independent fields, they are both determined 
by $\phi$. The fourth order free action for $\phi$ is positive definite, 
thus the path integral over all $\phi$ in Euclidean space
should converge, and should define a well determined Euclidean quantum field 
theory. 

One way to compute the path integral for a fourth order theory, is to 
expand $\phi$ in eigenfunctions of the differential operator $\hat O$
in the action. 
One then integrates over the coefficients in the harmonic expansion,
which  gives $ (\det \hat O )^{-1/2}$.
Another way is to use time slicing, by dividing 
the period into a number of short time steps $\epsilon$ and
approximating the derivatives by
\be
\phi_{,\tau} \sim \frac{(\phi_{n+1} - \phi_{n})}{\epsilon}
\quad
,
\quad
\phi_{,\tau \tau} \sim \frac{(\phi_{n+2} -2 \phi_{n+1} +\phi_{n})}{\epsilon^2}
\ee
One then integrates over the values of $\phi$ on 
each time slice. 
In a second order theory, where the action depends on $\phi$ 
and $\phi_{,\t}$ but not on $\phi_{,\t \t}$, 
the path integral will depend on the 
values of $\phi$ on the initial and final surfaces. However, in a fourth order 
theory, the use of three neighbor differences means that one has to specify 
$\phi_{,\t}$ on the initial and final surfaces as well. 

One can also see what needs to be specified on the initial and final
surfaces as follows. In classical second order theory, a state can be 
defined by its Cauchy data on a spacelike surface, i.e. the values of 
$\phi$ and $\phi_{,\tau}$ on the surface. 
In a canonical 3+1 treatment, these are regarded as the position of the 
field and its conjugate momentum.
In quantum theory, position and momentum don't commute, so instead 
one describes a state by a wave function in either position space 
or momentum space. In ordinary 
quantum mechanics, the position and momentum representations are regarded as 
equivalent: one is just the Fourier transform of the other. 
However, with path integrals, one has to use wave functions in the position 
representation.
This can be seen as follows. Imagine using the path 
integral to go from a state at $\tau_1$ to a state at $\tau_2$, 
and then to a state at $\tau_3$. 
In the position representation, the amplitude to go from a field $\phi_1$  
on $\tau_1$, to $\phi_2$ at $\tau_2$, is given by a path integral over all 
fields $\phi$ with the given boundary values. 
Similarly, the amplitude to go from $\phi_2$ 
at $\tau_2$, to $\phi_3$ at $\tau_3$, is given by another path integral. These 
amplitudes obey a composition law,

\be
G (\phi_3, \phi_1 ) = \int d \phi_2 
G (\phi_3, \phi_2 ) G (\phi_2, \phi_1  )
\ee

The composition law holds, only because one can join a field from $\phi_1$ to 
$\phi_2$ to a field from $\phi_2$ to $\phi_3$, 
to obtain a field from $\phi_1$ to 
$\phi_3$. Although in general $\phi_{,\t}$ will be discontinuous at $t_2$, the 
field will still have a well defined action, 
\be
S(\phi_3,\phi_1 ) = S(\phi_3,\phi_2 ) + S(\phi_2, \phi_3)
\ee
On the other hand, 
if one would use the momentum representation and wave functions in terms of 
$\phi_{,\t}$, the composition law would no longer hold, because 
the discontinuity of $\phi$ at $\tau_2$ would make the action infinite. 
Thus in second order theories, one should use wave functions in terms 
of $\phi$ rather than $\phi_{,\t}$. 

In a fourth order theory, a classical state is 
determined by the values of $\phi$ and its first three time derivatives on a 
spacelike surface. In a canonical treatment, $\phi$ and $\phi_{, \t \t}$ 
are usually taken to be independent coordinates. For the scalar field theory
(\ref{fourth}) we then have the conjugate momenta
\be
\Pi_{\phi} = -\phi_{,\t \t \t}+(m_{1}^{2} + m_{2}^{2} -2 \vec \nabla^2 ) 
\phi_{,\t}
\quad
,
\quad
\Pi_{\phi_{,\t \t}} = -\phi_{,\t}
\ee
This suggests that in quantum theory, one should describe a state by a 
wave functional $\Psi (\phi,\phi_{,\t \t})$ on a surface. Indeed, this is 
closely related to using the fields $\psi_1$ and $\psi_2$ that we introduced 
earlier. These were linear combinations of $\phi$ and $\Box \phi$,
thus taking the wave function to depend on $\psi_1$ and $\psi_2$, 
is equivalent to it depending on $\phi$ and $\phi_{,\t \t}$.
However, if one does the path 
integral between fixed values of $\phi$ and $ \phi_{, \t \t}$, one gets in 
trouble with the composition law, because the values 
of $\phi_{,\t}$ on the intermediate surface at $\t_2$ are not constrained, 
Hence $\phi_{,\t}$ will be in general discontinuous at $\t_2$, which implies
that $\phi_{,\t \t}$ will have a delta-function 
when one joins the fields above and below $\t_2$. 
In a second order action
$\phi_{,\t \t}$ appears linearly, thus the delta-function can be integrated by 
parts and the action of the combined field is finite. 
But in a fourth order action $( \phi_{,\t \t})^2$ appears, rendering
the action of the combined field infinite if $\phi_{,\t \t}$ 
is a delta-function.

Therefore, the path integral requires that quantum states 
be specified by $\phi$ and $\phi_{,\t} $ in order to get the composition law 
for amplitudes in a fourth order theory. 
In the next section we show how one can obtain transition
probabilities for observations
from the Euclidean path integral over $\phi$.

\section{Higher Derivative Harmonic Oscillator}

\subsection{Ground State Wave Function}

To illustrate how probabilities can be calculated, we consider a
higher derivative harmonic oscillator, for which in Euclidean signature
we take the action
\be \label{theory}
S = \int d \tau \left[ \frac{\alpha^2}{2} \phi^2_{,\t \t} + 
\frac{1}{2}\phi^2_{,\t}  + \frac{1}{2}m^2 \phi^2
\right] \ee
For $\alpha^2 > 0$, this is very 
similar to our scalar field model, since in the latter we can take Fourier
components so that spatial derivatives behave like masses.
The general solution to the equation of motion is given by
\be
\phi (\tau ) = 
A \sinh \lambda_1 \tau+ B\cosh \lambda_1 \tau+C \sinh \lambda_2 \tau+
D \cosh \lambda_2 \tau,
\ee
where $\lambda_1$ and 
$\lambda_2$ are given by (\ref{lambda}).
For small $\alpha$, $\lambda_1 \sim m$ and $\lambda_2 \sim 1/\alpha$.

The fourth order action for $\phi$ is positive definite, thus it gives a well 
defined Euclidean quantum field theory. In this theory, one can calculate the 
amplitude to go from a state $(\phi_1, \phi_{1,\t})$ at time $\tau_1$, 
to a state $(\phi_2,\phi_{2,\t})$ at time $\tau_2$. 
In particular, one can calculate the ground 
state wave function, the amplitude to go from zero field in the infinite 
Euclidean past, up to the given values $(\phi_0,\phi_{0,\t})$ at $\tau=0$.
This yields (see Appendix A)


\be\label{gr}
\Psi_0 (\phi_0,\phi_{0,\t} ) = N' \exp \left[-F'
\left( \phi_{0, \t}^2 + \frac{m}{\alpha} \phi_0^2 \right)
+\frac{2m^2 - m/\alpha}{(\lambda_2 -\lambda_1 )^2} \phi_0 \phi_{0,\t}\right]
\ee
where
\be
F'=\frac{(1 -4m^2 \alpha^2)}{2 \alpha^2
(\lambda_1 + \lambda_2 )(\lambda_2 -\lambda_1 )^2}
\ee
and $N'(\alpha,m)$ is a normalization factor.

Similarly, one can calculate the Euclidean conjugate ground state 
wave function $\Psi_0^{*}$, 
the amplitude to go from the given values at $\tau=0$, to zero 
field in the infinite Euclidean future. This conjugate wave function is 
equal to the original ground state wave function, with the opposite sign of 
$\phi_{0,\t}$.
The probability that a quantum fluctuation in the ground state gives the 
specified values $\phi_0$ and $\phi_{0,\t}$ on the surface $\tau =0$, 
is then given by


\be
P (\phi_0,\phi_{0,\t} ) = \Psi_0 \Psi_0^{*} =
N'^2 \exp \left[-2F'
\left( \phi_{0,\t}^2 + \frac{m}{\alpha} \phi_0^2 \right)\right]
\ee


The probability dies off at large values of $\phi$ and $\phi_{,\t}$
and is normalizable, thus the probability distribution in 
the Euclidean theory is well-defined. However if one Wick rotates to Minkowski
space, $\phi^2_{,\t}$ picks up a minus sign. The probability distribution 
becomes unbounded for large Lorentzian $\phi_{,\t}$ and can no longer 
be normalized. This 
is another reflection of the same problem as the ghosts. You can't fully 
determine a state on a spacelike surface, because that would involve 
specifying $\phi$ and Lorentzian $\phi_{,t}$, which
doesn't have a physically reasonable probability distribution.

Although one can not define a probability distribution for $\phi$ and 
Lorentzian $\phi_{,t}$ on a spacelike surface, one can calculate a 
probability distribution for $\phi$ alone, by integrating out over Euclidean 
$\phi_{,\t}$. This integral converges because 
the probability distribution is damped at large values of 
Euclidean $\phi_{,\t}$. This is just what one would 
calculate in a second order theory. So the moral is, a fourth order theory
can make sense in Lorentzian space, if you treat it like a second order theory.
The normalized probability distribution 
that a ground state fluctuations gives the specified value $\phi_0$ 
on a spacelike surface is then given by,


\be \label{groundprob}
P (\phi_0 ) =
\left(\frac{2F'm}{\pi \alpha}\right)^{1/2}
\exp \left[- \frac{2mF'}{\alpha}\phi_0^2 \right]
\ee

As the coefficient $\alpha$ of the fourth order term in the action tends to 
zero, this becomes
\be
P (\phi_0 ) = \left(\frac{m}{\pi}\right)^{1/2} (1+\frac{m\alpha}{2})
\exp [-m(1+m\alpha)\phi_0^2],
\ee
which tends toward the result for the second order theory.

\subsection{Transition Probabilities}

In this section we compute the Euclidean transition probability, to
go from a specified value $\phi_1$ at time $\tau_1$, to $\phi_2$ at time 
$\tau_2$, for the higher derivative harmonic oscillator.

In a second order theory, a state can be described by a wave function that 
depends on the values of $\phi$ on a spacelike surface. Thus 
a transition amplitude is given by a path integral from 
an initial state $\phi_1$ 
on $\tau_1$, to a final state $\phi_2$ on $\tau_2$. 
To calculate the probability to go from the initial state to 
the final, one multiplies the amplitude by its Euclidean conjugate. This can 
be represented as the path integral from a third surface, at $\tau_3$, back 
to $\tau_2$. 
Because the path integral in a second order theory depends only on 
$\phi$ on the boundary, what happens above $\tau_3$ and below $\tau_1$ 
doesn't matter. 
Furthermore, the path integrals above and below $\tau_2$ can be calculated 
independently, which implies the probability to go from initial to final, 
can be 
factorized into the product of an S matrix and its adjoint. The S matrix is 
unitary, because probability is conserved.

Now let us calculate the probability to go from an initial to a
final state in the fourth order theory (\ref{theory}). 
The path integral requires quantum states to be specified by
$\phi$ and $\phi_{,\t}$. The transition amplitude to go from a state 
$( \phi_1, \phi_{1,\t})$ at time $\tau_1=-T$, 
to a state $(\phi_2,\phi_{2,\t})$ at time $\tau_2=0$, reads

\be
\langle (\phi_2, \phi_{2,\t}; 0 ) \vert (\phi_1, \phi_{1,\t};-T) \rangle
 = \int_{(\phi_1,\phi_{1,\t})}^{(\phi_2,\phi_{2,\t})} 
d[\phi (\tau )] \exp [-S(\phi)]
\ee
This is evaluated in Appendix A, by writing $\phi = \phi_{cl} + \phi'$,
where $\phi_{cl}$ obeys the equation of motion with
the given boundary conditions on both surfaces.
\newpage
The result is
\ba
\langle (\phi_2, \phi_{2,\t}; 0 ) \vert (\phi_1,  \phi_{1,\t};-T) \rangle & =& 
\left(-\frac{\alpha(1+\alpha N) H}{2\pi^2} \right)^{1/2} 
\exp \left[-E(\phi_1^2 + \phi_2^2 )-F(\phi_{1,\t}^2 + \phi_{2,\t}^2 )
\right. \nonumber\\
 & -& \left.G \phi_{1,\t} \phi_{2,\t}+H \phi_1 \phi_2 
-K(\phi_{2,\t} \phi_2 - \phi_{1,\t} \phi_1 ) -L(\phi_{2,\t} \phi_1 
- \phi_{1,\t} \phi_2 )\right ]
\ea
The coefficient functions in the exponent are given by (\ref{coef}),
and $N$ is a normalization factor.

Again, one can construct a three layer 'sandwich' to calculate the
probability to go from the initial state to the final. 
However, in contrast with the second order theory
the path integral now depends on both $\phi$ and $\phi_{,\t}$ 
on the boundaries. This has two important implications for the
calculation of the transition probability. 
Firstly, as we just 
showed, one can't observe Lorentzian $\phi_{,\t}$ because it has an unbounded 
Lorentzian probability distribution. 
Therefore one should take $\phi_{,\t}$ to be 
continuous on the surfaces and integrate over all values, fixing only the 
values of $\phi$ on the surfaces.
Because the path integrals above and 
below $\tau_2 =0$ both depend on $\phi_{2,\t}$, the 
probability $P(\phi_2,\phi_1)$ to observe the initial and final
specified values of $\phi$
does not factorize into an S matrix and its adjoint. 
Instead, there is loss of quantum coherence, because one can 
not observe all the information that characterizes the final state.

After multiplying by the Euclidean conjugate amplitude and
integrating out over $\phi_{2,\t}$ we obtain
\be
 -\frac{\alpha(1+\alpha N) H}{2\pi^2} \left(\frac{\pi}{2F}\right)^{1/2}
\exp \left[-2E(\phi_1^2 + \phi_2^2 )-2F\phi_{1,\t}^2 +2H \phi_1 \phi_2 
+\frac{G^2}{2F}\phi_{1,\t}^2\right]
\ee

Another consequence of the dependence of the path integral on $\phi_{,\t}$ is 
that what goes on outside the sandwich, now affects the result. The most 
natural choice, would be the vacuum state above $\tau_3 =T$ 
and below $\tau_1 =-T$.
In other words, one takes the path integral to be over all fields 
that have the given values on the three surfaces, and that go to zero in the 
infinite Euclidean future and past. This means that to obtain the
transition probability we also ought to multiply
by the appropriately normalized 
ground state wave function $\Psi_0 (\phi_1,\phi_{1,\t})$ and its
Euclidean conjugate.
The probability $P(\phi_2,\phi_1 )$ is then given by
\ba
P(\phi_2,\phi_1) & = &
\int d[\phi_{1,\t} ] \Psi_0 \Psi_0^{*} \int d [\phi_{2,\t} ]
\langle (\phi_1, \phi_{1,\t}) \vert (\phi_2, \phi_{2,\t}) \rangle
\langle (\phi_2, \phi_{2,\t}) \vert (\phi_1,  \phi_{1,\t}) \rangle\nonumber\\
& = &  \left(\frac{\alpha^2(1+\alpha \tilde N)^2 H^2}{2\pi^2 (4F(F' +F) -
G^2)}\right)^{1/2}
\exp \left[ -2E (\phi_1^2 + \phi_2^2 )-2\frac{m F'}{\alpha}\phi_1^2
+2H \phi_1 \phi_2 \right]
\ea  
Here $F'(\alpha,m)$ is the coefficient in the exponent of 
the ground wave function (\ref{gr}) and $\tilde N$ is
a normalization factor.
In the limit $\alpha \rightarrow 0$, this reduces to
\be
P(\phi_2,\phi_1) = 
\frac{m}{2\pi \sinh m T}\exp \left[ -\frac{
m \cosh mT (\phi_1^2 +\phi_2^2 ) - 2m \phi_1 \phi_2 }{\sinh mT} 
-m \phi_1^2 \right]
\ee

Hence the 
probability given by the sandwich tends toward that of the second order 
theory, as the coefficient of the fourth order term in the action tends to 
zero. This is important, because it means that fourth order corrections to 
graviton scattering can be neglected completely at the low energies that 
now occur in the universe. 
On the other hand, in the very early universe, when fourth order terms are 
important, we expect the Euclidean metric to be some instanton, 
like a four sphere. In such a situation, one can not define 
scattering or ask about unitarity. The only quantities we have any chance of 
observing are the n-point functions of the metric perturbations, which 
determine the n-point functions of fluctuations in the microwave background. 
With Reall we have shown that Starobinsky's model of inflation 
\cite{Starobinsky80}, 
in which inflation is driven by the trace anomaly of 
a large number of conformally coupled 
matter fields, can give a sensible spectrum of microwave fluctuations, 
despite the fact it has fourth order terms and ghosts\cite{Hawking01}. 
Moreover, the fourth order terms can play an
important role in reducing the fluctuations to the level we observe.

Finally, in order to obtain the Minkowski space probability,
one analytically continues $\tau_2$ to future infinity in Minkowski space, 
and $\tau_1$ and 
$\tau_3$ to past infinity, keeping their Euclidean time values fixed. This gives 
the Minkowski space probability, to go from an initial value $\phi_1$
to a final value $\phi_2$.

\section{Runaways and Causality}


The discussion in Section II suggests that even the slightest amount of a 
fourth order term will lead to runaway 
production of positive and negative energies, or of real and ghost particles.
The classical theory is certainly unstable, if one prescribes the initial 
value of $\phi$ and its first three time derivatives. 
However, in quantum theory every sensible question can be posed in terms of 
vacuum to vacuum amplitudes. These can be defined by Wick rotating to 
Euclidean space and doing a path integral over all fields that die off in the 
Euclidean future and past. Thus the Euclidean formulation of a quantum field 
theory implicitly imposes the final boundary condition that the fields remain 
bounded. This removes the instabilities and runaways, like a final boundary 
condition removes the runaway solution of the classical radiation reaction 
force. The price one pays for removing runaways with a final boundary 
condition, is a slight violation of causality. 
For instance, with the classical radiation reaction force, 
a particle would start to accelerate before a wave hit it. 
This can be seen by considering a single electron which is acted upon
by a delta-function pulse \cite{Coleman69}. The equation of motion 
for the $x$-component reduces to
\be
x_{,tt} = \lambda x_{,ttt} +\delta (t),
\ee
with $\lambda =\frac{2e^2}{3mc^3}$. This has the solution
\be
x(t) = \int \frac{d\omega}{2\pi} \exp [ -i\omega t]
\frac{1}{-\omega^2 -i\lambda \omega^3}.
\ee
The integrand has two singularities, at $\omega =0$ and 
$\omega = i\lambda^{-1}$. The final boundary condition that $x_{,t}$
should tend to a finite limit, implies
one must choose an integration contour that stays close 
to the real axis, going below the second singularity.
This yields
\ba
x(t) & = & \lambda \exp [t/\lambda ], \quad t<0 \nonumber\\
& = & t +\lambda, \quad \ \ \ \ \quad t>0
\ea
which is without runaways, but acausal.
\newpage
However, this pre-acceleration is appreciable only
for a period of time comparable with the time for light to
travel the classical radius of the electron, and thus
practically unobservable.

Similarly, if we would add an interaction term to the higher derivative 
scalar field theory (\ref{fourth}), the
imposition of a final boundary condition to eliminate the runaway solutions,
would lead to acausal behaviour on the scale of $m_2^{-1}$,
where $m_2$ is the mass of the ghost particle.
However, in the context of quantum gravity,
one could again never detect a violation of causality, 
because the presence of a mass introduces a logarithmic time delay
$\Delta t \sim -m \log b$, where $b$ is the impact parameter.
Thus there is no standard arrival time, one can always arrive before any 
given light ray by taking a path which stays a sufficiently large distance
from the mass.

\section{Concluding Remarks}

We conclude that quantum gravity with fourth order corrections
can make sense, despite apparently having negative energy solutions and 
ghosts. In doing this, we seem to go against the convictions of the 
last 25 years,  that unitarity and causality are essential requirements of 
any viable theory of quantum gravity. Perturbative string theory has 
unitarity and causality, so it has been claimed as the only viable quantum 
theory of gravity. 
But the string perturbation expansion 
does not converge, and string theory has to be augmented by non perturbative 
objects, like D-branes. One can have a world-sheet theory of 
strings without higher derivatives, only because two dimensional metrics 
are conformally flat, meaning perturbations don't change the light cone.
Still, we live either on a 3-brane, 
or in the bulk of a higher dimensional compactified space.  
The world-sheet theory of D-branes with $p$ greater 
than one has similar non-renormalizability problems to Einstein gravity and 
supergravity. Thus string theory effectively has ghosts, though this awkward 
fact is quietly glided over.

To summarize, we showed that perturbation theory for gravity 
in dimensions greater than two required higher derivatives in the free 
action. Higher derivatives seemed to lead to ghosts, states with negative 
norm. To analyze what was happening, we considered a fourth order 
scalar field theory. We showed that the problem with ghosts arises because 
in the canonical approach, $\phi$ and $\Box \phi$ are regarded as two 
independent coordinates. Instead, we based quantum theory on a path 
integral over $\phi$, evaluated in Euclidean space and then Wick rotated 
to Lorentzian space. We showed the path integral required that quantum states 
be specified by the values of $\phi$ and $\phi_{,\t}$ on a spacelike surface, 
rather than $\phi$ and $\phi_{,\t \t}$ as is usually done in a canonical 
treatment. The wave function in terms of $\phi$ and $\phi_{,\t}$ 
is bounded in Euclidean space, but grows exponentially with Minkowski space 
$\phi_{,\t}$. This means one can not observe $\phi_{,\t}$ but only $\phi$. 
To calculate probabilities for observations one therefore has to trace out 
over $\phi_{,\t}$ on the final surface,
and lose information about the quantum state. 
One might worry that integrating out $\phi_{,\t}$ would break Lorentz 
invariance. However, $\phi_{,\t}$ is conjugate to $\phi_{,\t \t}$ so tracing 
over $\phi_{,\t}$ is equivalent to not observing $\Box \phi$. Since,
according to eq.(\ref{ps}),
$\psi_1$ and $\psi_2$ are linear combinations of $\phi$ and $\Box \phi$, this 
means that one only considers Feynman diagrams whose external legs are 
$\psi_1 -\psi_2$. You don't observe the other linear combination, 
$m_{2}^{2} \psi_1  -m_{1}^{2} \psi_2$.

Because one is throwing away information, 
one gets a density matrix for the final state, and loses unitarity. 
However, one can never produce a negative norm 
state or get a negative probability. 
We illustrated with the example of a higher derivative harmonic 
oscillator that probabilities for observations tend toward those of the
second order theory, as the 
coefficient of the fourth order term in the action tends to zero.
This means that the departures from unitarity for higher derivative gravity
will be 
very small at the low energies that now occur in the universe. 
On the other hand, the higher derivative terms 
will be important in the early universe, but there unitarity can not be 
defined. 

\bigskip
\centerline{\bf Acknowledgements}

It is a pleasure to thank David Gross, Jim Hartle and Edward Witten
for helpful discussions. We would also like to thank the ITP at Santa Barbara
and the Department of Physics at Caltech, where some of this work was done, 
for their hospitality. TH is Aspirant FWO, Belgium.

\appendix

\section{Transition Amplitude}

We compute the Euclidean transition amplitude, 
to go from an initial state $(\phi_1, \phi_{1,\t} )$ 
on a spacelike surface at $\tau =-T$,
to a final state  $(\phi_2, \phi_{2,\t})$ at $\tau = 0$, 
for the higher derivative harmonic oscillator (\ref{theory}).
The general solution to the equation of motion is given by
\be
\phi (\tau ) = 
A \sinh \lambda_1 \tau+ B\cosh \lambda_1 \tau+C \sinh \lambda_2 \tau+
D \cosh \lambda_2 \tau,
\ee
where 
\be \label{lambda}
\lambda_1 = \frac{1}{\sqrt{2\alpha^2}}\sqrt{(1-\sqrt{1-4m^2 \alpha^2})}
\quad
,
\quad 
\lambda_2 = \frac{1}{\sqrt{2\alpha^2}}\sqrt{(1+\sqrt{1-4m^2 \alpha^2})}
\ee

The transition amplitude is given by a path integral,
\be
\langle (\phi_2, \phi_{2,\t}; 0) \vert (\phi_1, \phi_{1,\t};-T) \rangle
 = \int_{(\phi_1,\phi_{1,\t})}^{(\phi_2,\phi_{2,\t})} 
d[\phi (\tau )] \exp [-S(\phi)]
\ee
This can be evaluated by separating out the 'classical' part of 
$\phi$. 
If we write $\phi = \phi_{cl} + \phi'$,
where $\phi_{cl}$ obeys the equation of motion with
the required boundary conditions on both surfaces $\tau=0$ and $\tau=T$, 
then the amplitude becomes
\be \label{ampl}
\langle (\phi_2, \phi_{2,\t}; 0 ) \vert (\phi_1, \phi_{1,\t};-T) \rangle  = 
\exp [-S_{cl} (\phi_1, \phi_{1,\t}, \phi_2, \phi_{2,\t} )]
\int_{(0,-T)}^{(0,0)} d[\phi' (\tau )] \exp [-S(\phi')]
\ee
The classical action is
\ba
S_{cl} & = &
\int_{0}^{T} d\tau \left[\frac{\alpha^2}{2} \phi_{cl,\t \t}^2 + 
\frac{1}{2}\phi_{cl,\t}^2  + \frac{1}{2}m^2 \phi_{cl}^2 \right]\nonumber\\
& = & E(\phi_1^2 + \phi_2^2 )+F(\phi_{1,\t}^2 + \phi_{2,\t}^2 )
+G \phi_{1,\t} \phi_{2,\t} -H \phi_1 \phi_2 \nonumber\\
& & 
\qquad +K(\phi_{2,\t} \phi_2 - \phi_{1,\t} \phi_1 ) +L(\phi_{2,\t} \phi_1 
-  \phi_{1,\t} \phi_2 )
\ea
where
\ba \label{coef}
E & = & \frac{ -m(1-4m^2\alpha^2)}
{2\alpha^3(\lambda_2^2 - \lambda_1^2)P^2} \left[
\frac{2m}{\alpha} (\cosh \lambda_2 T - \cosh \lambda_1 T)
(\lambda_1 \sinh \lambda_1 T +\lambda_2 \sinh \lambda_2 T )\right.
\nonumber\\
& & \left.
+ \sinh \lambda_1 T \sinh \lambda_2 T (\lambda_1(2\lambda_2^2 +\frac{1}
{\alpha^2})
\cosh \lambda_2 T \sinh \lambda_1 T-\lambda_2(2 \lambda_1^2 +
\frac{1}{\alpha^2})\sinh \lambda_2 T\cosh \lambda_1 T) \right]\nonumber\\
F & = & \frac{ (1-4m^2\alpha^2)}
{2\alpha^2(\lambda_2^2 - \lambda_1^2)P^2} \left[
\frac{2m}{\alpha}(\cosh \lambda_2 T - \cosh \lambda_1 T)
(\lambda_2 \sinh \lambda_1 T +\lambda_1 \sinh \lambda_2 T )\right.
\nonumber\\
& & \left.
+ \sinh \lambda_1 T \sinh \lambda_2 T (\lambda_2(2\lambda_1^2 +\frac{1}
{\alpha^2})
\cosh \lambda_2 T \sinh \lambda_1 T-\lambda_1(2 \lambda_2^2 +
\frac{1}{\alpha^2}) \sinh \lambda_2 T\cosh \lambda_1 T) \right]\nonumber\\
G & = & \frac{ (1-4m^2\alpha^2)}
{\alpha^2(\lambda_2^2 - \lambda_1^2)P^2} \left
(\frac{2m}{\alpha}(\cosh \lambda_2 T \cosh \lambda_1 T -1)\right.\nonumber\\
& & \left. \qquad \qquad \qquad \qquad
-\frac{1}{\alpha^2} \sinh \lambda_1 T \sinh \lambda_2 T \right)
(\lambda_1\sinh \lambda_2 T-\lambda_2\sinh \lambda_1 T )\nonumber\\
H & = &  \frac{ -m(1-4m^2\alpha^2)}
{\alpha^3(\lambda_2^2 - \lambda_1^2)P^2} \left
(\frac{2m}{\alpha}(\cosh \lambda_2 T \cosh \lambda_1 T -1)
\right. \nonumber\\
& & \left. \qquad \qquad  \qquad \qquad
-\frac{1}{\alpha^2}\sinh \lambda_1 T \sinh \lambda_2 T \right)
(\lambda_1\sinh \lambda_1 T-\lambda_2\sinh \lambda_2 T )\nonumber\\
K & = & 
 \frac{1}{P^2} \left[ 
\frac{m}{\alpha}\left(4m^2 +\frac{1}{\alpha^2} \right) 
\sinh \lambda_1 T \sinh \lambda_2 T
(1-\cosh \lambda_1 \cosh \lambda_2 T) \right. \nonumber\\
& & \qquad \qquad \qquad \qquad \left. + \frac{2m^2}{\alpha^2}
(2-3(\cosh^2 \lambda_1 T + \cosh^2 \lambda_2 T ))
\right] \nonumber\\
L & = &\frac{ -m(1-4m^2\alpha^2)}
{\alpha^3(\lambda_2^2 - \lambda_1^2)P^2} \left
(\frac{2m}{\alpha}(\cosh \lambda_2 T \cosh \lambda_1 T -1)\right.\nonumber\\
& & \qquad \qquad \qquad \qquad \left. - \frac{1}{\alpha^2}
\sinh \lambda_1 T \sinh \lambda_2 T\right)
(\cosh \lambda_2 T-\cosh \lambda_1 T )
\ea
with
\be
P= (\lambda_1^2 + \lambda_2^2 ) \sinh \lambda_1 T \sinh \lambda_2 T +2
\lambda_1 \lambda_2 (1- \cosh \lambda_1 T \cosh \lambda_2 T)
\ee

The pre-exponential factor in (\ref{ampl})
can be derived from the classical action alone
\cite{Marinov80}, it is basically the Jacobian of the 
change of variables $(\pi_1,\phi_1) \rightarrow (\phi_2,\phi_1)$.
Because the Lagrangian is quadratic, the prefactor is independent
of the values specifying the initial and final states,
and the transition amplitude (\ref{ampl}) is exact.
It is given by
\ba \label{am}
\langle (\phi_2, \phi_{2,\t}; 0) \vert (\phi_1, \phi_{1,\t};-T) \rangle
& =&  \left(\frac{-\alpha(1+\alpha N) H}{2\pi^2 }\right)^{1/2} 
\exp \left[-E(\phi_1^2 + \phi_2^2 )-F(\phi_{1,\t}^2 + \phi_{2,\t}^2 )
\right.\nonumber\\
& - & \left. G \phi_{1,\t} \phi_{2,\t} +H \phi_1 \phi_2 
 -K( \phi_{2,\t} \phi_2 - \phi_{1,\t} \phi_1 ) -L( \phi_{2,\t} \phi_1 
- \phi_{1,\t} \phi_2 )\right ]
\ea

The normalization factor $N$ is independent of $\alpha$ to first order. It 
is determined 
by taking $T \rightarrow +\infty $ in (\ref{am}) and requiring that 
the amplitude tends toward the product of two normalized ground state 
wave functions $\Psi_0 (\phi_1,\phi_{1,\t} )$ and 
$\Psi_0 (\phi_2, \phi_{2,\t} )$.

For small $\alpha$, $\lambda_1 \sim m$ and $\lambda_2 \sim 1/\alpha$,
hence the transition amplitude becomes
\ba
\langle (\phi_2, \phi_{2,\t}; 0 ) \vert (\phi_1,  \phi_{1,\t};-T) \rangle
& = &   
 \left(\frac{m\alpha}{2\pi^2 \sinh mT}\right)^{1/2} 
\exp \left[-\frac{m \cosh mT(\phi_1^2 + \phi_2^2 )}
{2 \sinh mT} -\frac{\alpha}{2}(\phi_{1,\t}^2 + \phi_{2,\t}^2 ) \right.
\nonumber\\
& & 
\left. 
+\frac{m\alpha \cosh mT (\phi_{2,\t} \phi_2 - \phi_{1,\t} \phi_1 )}{\sinh mT} 
-\frac{m(\alpha \phi_{2,\t} +\phi_2)(\alpha \phi_{1,\t} - \phi_1 )} 
{\sinh mT} \right ]
\ea


\begin{thebibliography}{99}

\bibitem{Besse87}A.L. Besse, {\em Einstein Manifolds}, 
Springer-Verlach, Berlin (1987), p.457.

\bibitem{thooft}G. 't Hooft and M. Veltman, Ann. Inst. Poincare 20, 69 (1974).

\bibitem{Stelle}K.S. Stelle, Phys. Rev. D16, 953 (1977).

\bibitem{Fradkin}E.S. Fradkin and A.A. Tseytlin, Nucl. Phys. B201, 469 (1982).

\bibitem{Hawking85}S.W. Hawking, in {\it Quantum Field Theory and Quantum 
Statistics:Essays in Honor of the 60th Birthday of E.S. Fradkin}, eds. 
A. Batalin, C.J. Isham and C.A. Vilkovisky, Hilger, Bristol, UK (1987).

\bibitem{Tomboulis77}E. Tomboulis, Phys. Lett. 70B, 361 (1977).

\bibitem{Johnston87}D.A. Johnston, Nucl. Phys. B297, 721 (1987).

\bibitem{Starobinsky80}A.A. Starobinsky, Phys. Lett. 91B, 99 (1980).

\bibitem{Hawking01}S.W. Hawking, T. Hertog, H.S. Reall,
Phys. Rev. D63, 083504 (2001).

\bibitem{Coleman69}S. Coleman in {\em Theory and Phenomenology in Particle 
Physics}, ed. A. Zichichi, New York (1969).

\bibitem{Marinov80}M.S. Marinov, Phys. Rep. 60, 1 (1980).





\end{thebibliography}
\end{document}